\documentstyle[preprint,aps,epsf]{revtex}
\def\rts{\sqrt s}

\def\mw{M_W}
\def\mz{M_Z}
\def\h{H}           
\def\mh{m_{\h}}
\def\mt{m_t}

\def\fbi{~{\rm fb}^{-1}}

\def\mev{~{\rm MeV}}
\def\gev{~{\rm GeV}}

\def\anti{\overline}

\def\eg{{\it e.g.}}
\def\etal{{\it et al.}}
\def\ie{{\it i.e.}}
\def\epem{{e^+e^-}}
\def\mupmum{{\mu^+\mu^-}}
\def\br{B}
\def\lsim{\alt}

\def\stop{\widetilde t}
\def\mstop{m_{\stop}}
\def\hl{h^0}
\def\mhl{m_{\hl}}

\begin{document}

\preprint{
\font\fortssbx=cmssbx10 scaled \magstep2
\hbox to \hsize{
\hfill$\vcenter{
             \hbox{\bf UCD-96-39}
             \hbox{\bf MADPH-96-979}
             \hbox{\bf IUHET-351}
             \hbox{December 1996}}$ }
}

\title{\vspace*{1in}
Precision Higgs boson mass determination\\ at lepton colliders}

\author{V. Barger$^a$, M.S.~Berger$^b$, J.F.~Gunion$^c$, and T.~Han$^c$}

\address{
$^a$Physics Department, University of Wisconsin, Madison, WI 53706, USA\\
$^b$Physics Department, Indiana University, Bloomington, IN 47405, USA\\
$^c$Physics Department, University of California, Davis, CA 95616, USA}

\maketitle

\thispagestyle{empty}

\begin{abstract}

We demonstrate that a measurement of the Bjorken process
$e^+e^-, \mu^+\mu^-\to ZH$ in the threshold region can yield a precise
determination of the Higgs boson mass. With
an integrated luminosity of $100\fbi$,
it is possible to measure the Higgs mass to within 60\mev\
(100\mev) for $\mh=100\gev$ ($150\gev$). 

\end{abstract}

\newpage

One of the triumphs of the LEP program was the measurement of the
$Z$-boson mass to two MeV. Expectations are also quite good for the
measurement of the $W$-boson mass ($\mw$) 
and the top quark mass ($\mt$) in the future, perhaps achieving
precision of order 10~MeV for $\mw$ and $2\gev$
for $\mt$ at the Tevatron and the LHC~\cite{snowmassmw}.
Precise values for $\mw$ and $\mt$ 
can also be obtained at lepton colliders by measuring the 
$\ell^+\ell^-\to WW$ and $\ell^+\ell^-\to t\overline{t}$ 
($\ell = e$ or $\mu$) threshold cross sections, as illustrated
by measurements of $W$-pair production at LEP
center-of-mass energy $\rts=161\gev$ \cite{lep161}.
These measurements will allow an indirect prediction for the Higgs
boson mass ($\mh$) and will test the
consistency of the Standard Model (SM) at the two-loop level
once $\mh$ is known.

In this Letter we point out that, analogously,
a very accurate determination of $\mh$ is obtained
by measuring the threshold cross section for the Bjorken Higgs-strahlung
process \cite{bj} $\ell^+\ell^-\to Z\h$;
with integrated luminosity $L=100\fbi$, a
$1\sigma$ precision of order 60 MeV is possible for $\mh=100\gev$.
This error in $\mh$ is smaller than that achievable 
via final state mass reconstruction for a typical detector, and would then
be the most accurate determination of $\mh$ at an $\epem$ collider.

The SM Higgs boson is easily discovered in the $Z\h$ production
mode by running the machine well above threshold, \eg\ at $\rts=500\gev$.
For $\mh\lsim 2\mw$ the dominant Higgs boson decay is to $b\anti b$
and most backgrounds can be eliminated by $b$-tagging.
At the next linear $\epem$ collider (NLC)
the accuracy for $\mh$ via reconstruction using final state momenta
is strongly dependent on the detector performance and signal statistics:
$\Delta \mh \simeq R_{\rm event}(\gev)/\sqrt N $,
where $R_{\rm event}$ is the single-event resolution and $N$ is the number
of signal events. At an SLD-type detector,
the single event resolution for reconstruction
of the Higgs mass is about $4\gev$ for most $Z\h$ final states
(including channels with $Z\to\epem,\mupmum$) \cite{janot}.
At the ``super''-LC detector \cite{jlci},
the Higgs mass measurement would be best performed by examining
the mass spectrum of the system recoiling against $Z\to\epem,\mupmum$ decays.
The resolution in this spectrum would be about $0.3\gev$ \cite{jlci,kawagoe}.
For the SM Higgs boson, the accuracies of the $\mh$ determination
for the two types of detector are\footnote{The LHC collaborations
expect that the SM Higgs boson is
detectable in the mass range $50\alt \mh\alt 150\gev$ via its
$\gamma \gamma$ decay mode.
The mass resolution is expected to be $\alt 1\%$.}
\begin{eqnarray}
{\rm SLD}: \Delta \mh\simeq 180\mev\left ({{50\fbi}\over {L}}\right )^{1/2}\,,
{}~~~ {\rm super-LC}:
\Delta \mh\simeq 20\mev\left ({{50\fbi}\over {L}}\right )^{1/2}\,,
\label{sldjlcerrors}
\end{eqnarray}
which take into account the effective branching 
ratios appropriate in the two different cases.
The super-LC accuracy would be competitive with that we shall obtain
via the threshold technique. However, in the not unlikely case
that the detector is of the SLD-type, the best means for measuring $\mh$
will be to first determine $\mh$ to within a few hundred MeV in
$\rts=500\gev$ running [which will also yield a precise
measurement of $\sigma(Z\h)$] and then reconfigure the collider
for maximal luminosity just above the threshold energy $\rts= \mz+\mh$.

\begin{figure}[htb]
\leavevmode
\begin{center}
\epsfxsize=4.0in\hspace{0in}\epsffile{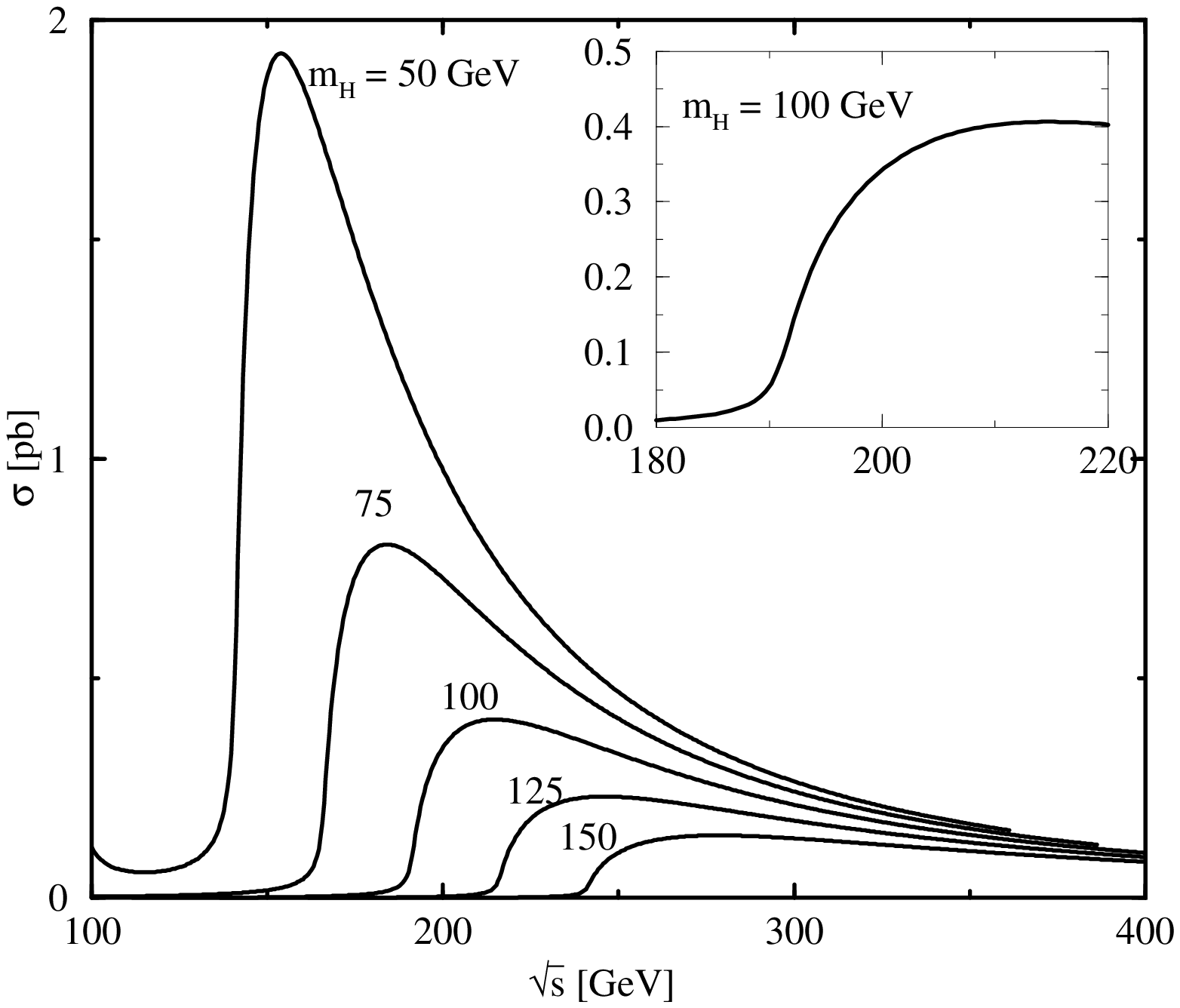}
\end{center}
\caption[]{\footnotesize\sf The cross section vs. $\protect\rts$
for the process
$\ell^+\ell^-\to Z^\star\h\to f\anti f\h$ for a range of Higgs masses. 
The inset figure shows the
detailed structure for $\mh=100$~GeV in the threshold region.}
\label{zhcurve}
\end{figure}

In Fig.~\ref{zhcurve} we show the cross section for the Bjorken process
$\ell^+\ell^-\to Z\h$ for Higgs masses from 50 to 150~GeV. Since the threshold
behavior is $S$-wave, the rise in the cross section in the threshold region
is rapid, as can be seen for the case of
$\mh=100$~GeV in the inset figure, the cross section being
a few tenths of a pb. 
At LEP II, the few hundred pb$^{-1}$ of luminosity that might
be devoted to such a threshold would yield just a handful of events.
However,
much higher luminosity is possible at threshold at the NLC \cite{NLCo}
or a muon collider \cite{bbgh1,bbgh2,feas}.

In the ideal case that the normalization of the measured $Z\h$ cross section
as a function of $\rts$ can be precisely predicted, including
efficiencies and systematic effects,
sensitivity to the SM Higgs boson mass is maximized by
a single measurement of the cross section at $\rts=\mz+\mh+0.5$~GeV,
just above the real particle threshold. With a $\sim\pm 180\mev$
measurement of $\mh$ from initial running
[see Eq.~(\ref{sldjlcerrors})] $\rts$
can be set quite close to this optimal point.
As an example of the precision that might be achieved, suppose $\mh=100$~GeV
and backgrounds are neglected.
The $Z\h$ cross section is 120~fb and is rising at a rate of 0.05~fb/MeV.
With $L=50\fbi$ and including an overall
($b$-tagging, geometric and event identification)
efficiency of $40\%$,
this yields $2.4\times 10^3$ events, or a measurement of the
cross section to about 2\%. From the slope of the cross section one concludes
that a $\mh$ measurement with accuracy of roughly 50~MeV is possible.

In practice, there will be systematic errors associated with
experimental efficiencies as well as for theoretical predictions of the
$Z\h$ cross section and $\h$ branching ratio(s) that will
be very difficult to reduce below the 1\% level.
The ratio of the cross section measured
at $\rts$ well-above threshold in the initial $H$ discovery
to that measured right at threshold is thus the key to determining $\mh$.
The theoretical uncertainties will cancel in the ratio. Given
the high luminosity that should be available for measurements
both well-above threshold and right at threshold,
changes in $b$-tagging, geometrical efficiencies,
and jet misidentification as a function of $\rts$ may be understood at the
$<1\%$ level, provided
the final-focus reconfiguration required to optimize luminosity
at the lower threshold $\rts$ does not impel detector changes that
would lead to significant changes in the experimental systematic effects.

\begin{figure}[htb]
\leavevmode
\begin{center}
\epsfxsize=3.5in\hspace{0in}\epsffile{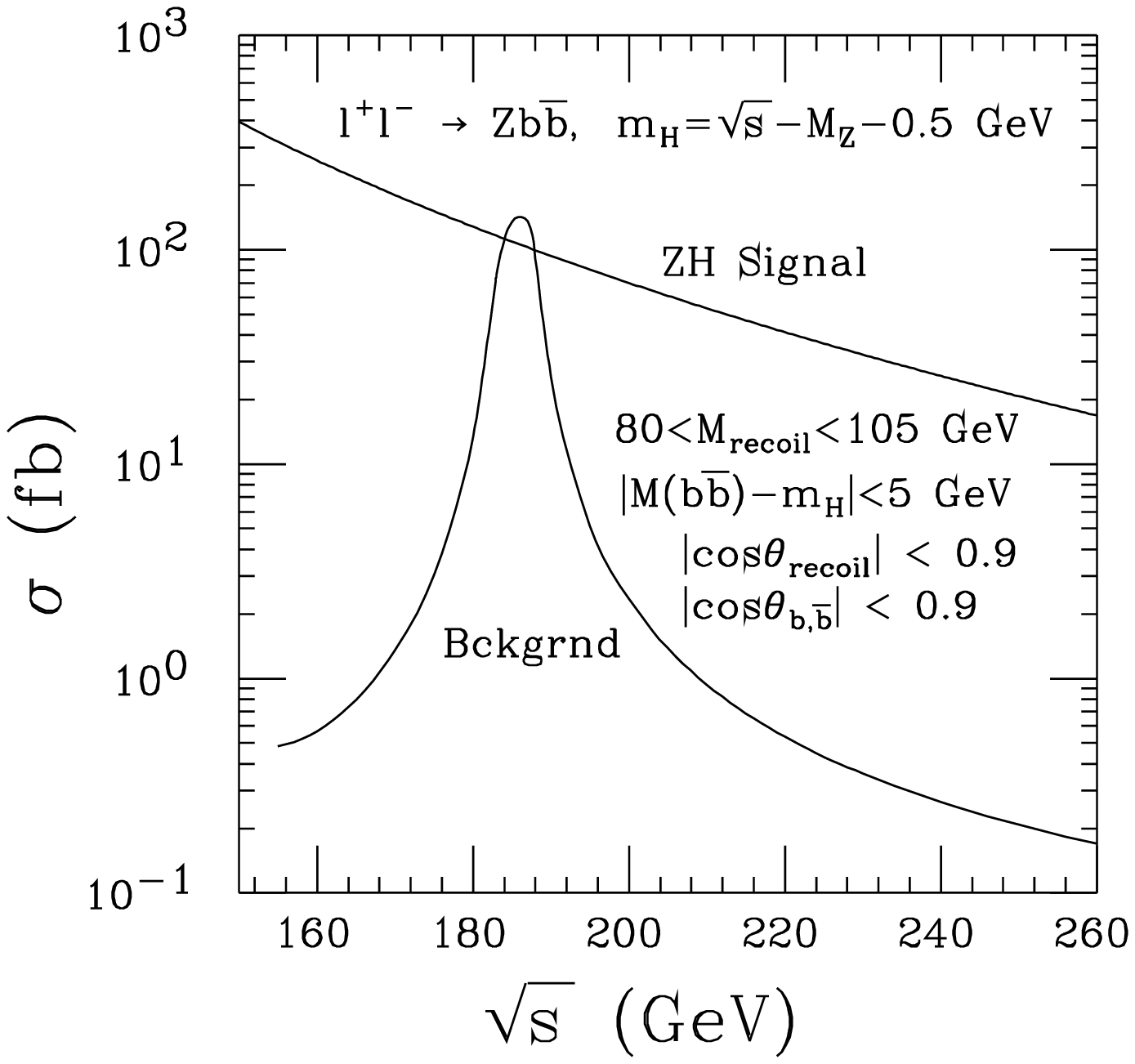}
\end{center}
\caption[]{\footnotesize\sf The $\ell^+\ell^-\to Z\h\to Z b\anti b$
signal and the irreducible $\ell^+\ell^-\to Zb\overline{b}$ background
vs. $\rts$, including $b$-tagging and cut requirements 1)-4), see text.}
\label{eezbb}
\end{figure}

For a more precise estimate of the accuracy
with which $\mh$ can be measured, we employ $b$-tagging and cuts in order to
reduce the background to a very low level. Specifically,
we require: 1) tagging of both $b$'s in the event (for which an overall
50\% efficiency will be assumed);
2) $|M_{b\anti b}-\mh|<5\gev$;
3) $80<M_{\rm recoil}<105\gev$
(\ie\ broadly consistent\footnote{Note that the restriction
on $M_{\rm recoil}$
means that constructive interference of $Z\h$
diagrams with $WW$ ($ZZ$) fusion diagrams
in the $\nu_{\ell}\overline\nu_{\ell}\h$ ($\ell^+\ell^-\h$)
channels \cite{bsss} will be small.}
with $\mz$),
where $M_{\rm recoil}\equiv[p_{\rm recoil}^2]^{1/2}$ with
$p_{\rm recoil}= p^{}_{\ell^+}+p^{}_{\ell^-}-p^{}_b-p^{}_{\anti b}$\,;
4) $|\cos\theta_{b,\anti b,{\rm recoil}}|<0.9$, where $\theta$
is the polar angle with respect to the beam direction.
With these cuts the only significant background
will be that from $ZZ$ production, where at least one of the $Z$'s
decays to $b\anti b$. In Fig.~\ref{eezbb}, we compare the cross section
versus $\rts$
for the $\ell^+\ell^-\to Zb\overline{b}$ background to that for the
$\ell^+\ell^-\to Z\h$ (with $\h\to b\anti b$) signal, where the signal
is computed for $\mh=\rts-\mz-0.5\gev$.
The background is very much smaller than the signal unless $\mh$
is close to $\mz$.

\begin{figure}[htb]
\leavevmode
\begin{center}
\epsfxsize=3.5in\hspace{0in}\epsffile{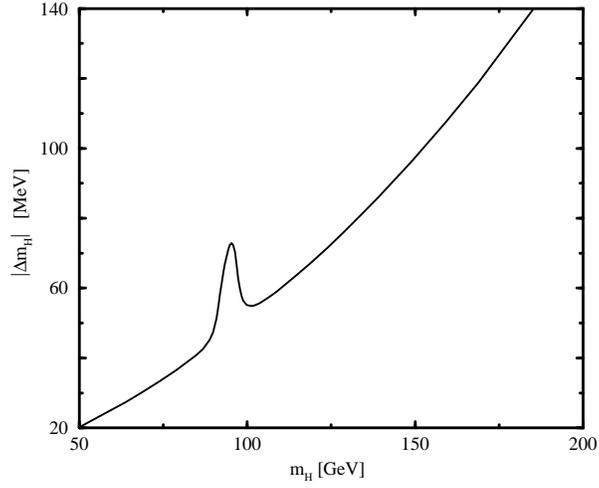}
\end{center}
\caption[]{\footnotesize\sf
The precision $\Delta \mh$ attainable from a $50\fbi$
measurement of the $Z b\anti b$ cross section at $\protect\rts=\mz+\mh+0.5\gev$
as a function of $\mh$, including $b$-tagging and cuts
1)-4). Bremsstrahlung, beamstrahlung, and beam energy smearing are neglected.
A precise measurement of the cross section well above threshold
is presumed available.}
\label{error}
\end{figure}

The expected precision for the Higgs mass is given in Fig.~\ref{error} for an
integrated luminosity of 50~fb$^{-1}$. The precision degrades
as $\mh$ increases
because the signal cross section is smaller (see Fig.~\ref{zhcurve}).
The background from the $Z$-peak reduces the precision
for $\mh\approx \mz$. Bremsstrahlung, beamstrahlung and
beam energy smearing yield a reduction in sensitivity of 15\%
at a muon collider and 35\% at an $e^+e^-$ collider. 

If the Higgs boson is discovered at the Tevatron or LHC prior to
construction of the NLC,
the NLC could be configured from the beginning for optimal luminosity
in the vicinity of the $Z\h$ threshold.\footnote{If instead
a muon collider is the first to be constructed following $\h$ discovery, 
then the appropriate first
emphasis might be $s$-channel $\h$ production, which allows extremely
accurate mass, width and coupling-constant-ratio determinations
\cite{bbgh1,bbgh2}.}
A motivation for doing so is that, at the NLC,
measurements near the peak in the $Z\h$ cross section (not far above
threshold) would yield the highest rates and, hence, smallest errors possible
for determining the branching ratios, couplings and total width of the $\h$.
In order to determine both these $\h$ properties and also
$\mh$, a very useful first set of measurements would be to take data
at $\rts=\mh+\mz+20\gev$ and at $\rts=\mh+\mz+0.5\gev$. In particular,
the $\epem\to Z\h\to Zb\anti b$ rates at these two energies
would simultaneously determine $\mh$ and $\sigma(Z\h)\br(\h\to b\anti b)$,
where $\sigma(Z\h)\propto g_{ZZ\h}^2$, the square of 
$ZZ\h$ coupling strength. The
inclusive (recoil spectrum) $Z\h$ event rate would yield a determination
of $\sigma(Z\h)$ directly and $\br(\h\to b\anti b)$ could then be computed;
deviations in either from SM expectations would be of great interest.

\begin{figure}[htb]
\leavevmode
\begin{center}
\epsfxsize=3.8in\hspace{0in}\epsffile{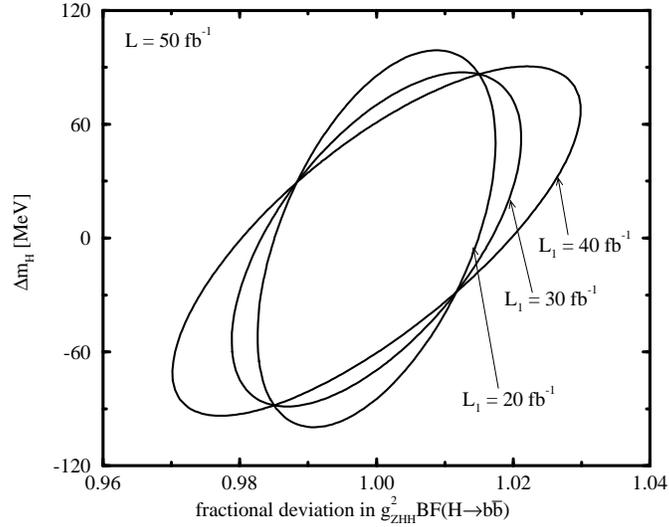}
\end{center}
\caption[]{\footnotesize\sf The $\Delta \chi ^2=1$ contours for determining
the Higgs mass and $g_{ZZ\h}^2\br(\h\to b\anti b)$ by devoting
$L_1+L_2=50\fbi$ to two points along the threshold curve: $L_1$ at
$\protect\rts=\mz+\mh+0.5$~GeV and $L_2$ at $\protect\rts=\mz+\mh+20$~GeV.
We have assumed $\mh=100\gev$; $b$-tagging and cuts 1)-4) are imposed.}
\label{zhsin2}
\end{figure}

Figure~\ref{zhsin2} shows the statistical precision that can be
obtained in a two parameter fit to $\mh$ and $g_{ZZ\h}^2\br(\h\to b\anti b)$
(before including smearing effects from bremsstrahlung, beamstrahlung
and beam energy spread) using a combined integrated luminosity of
50~fb$^{-1}$ for the above two values of $\rts$ in the threshold region.
The smallest error in $\mh$ is $\sim\pm 85\mev$ (for $\mh=100\gev$)
obtained with $L_1=30\fbi$ at $\rts=\mh+\mz+0.5$ and $L_2=20\fbi$ at
$\rts=\mz+\mh+20$~GeV.
Since $\mh$ is determined by the ratio of the cross sections
at the two energies, systematic uncertainties would cancel almost
completely for such closely spaced energies, and the error in $\mh$
would be almost entirely statistical.  The measurement of
$\sigma(Z\h)\br(\h\to b\anti b)$
would be at the $\pm 2\%$ statistical level (which
is better than the precision that can be reached with $L=200\fbi$
accumulated at $\rts=500\gev$ \cite{summary});
at this level of statistical
error, the systematic uncertainties on $\sigma \br$ 
from $b$-tagging, geometrical cuts
and event-identification efficiencies will probably dominate.
%
Doubling $L=L_1+L_2$ to $100\fbi$ (so that $L_1=60\fbi$)
would yield $\sim\pm60\mev$ error for $\mh$,  \ie\
comparable to the error of $\sim \pm 55\mev$ shown in Fig.~\ref{error}
for $L_1=50\fbi$ assuming small statistical error for 
$\sigma(Z\h)\br(\h\to b\anti b)$ at $\rts=500\gev$.

Two comments are particularly relevant. First,
a $\pm 60\mev$ uncertainty on $\mh$ would allow almost
immediate centering on the $s$-channel Higgs resonance peak at
a muon collider (thereby avoiding expending luminosity on a scan location
of the peak). For $\mh\lsim 2\mw$, 
a fine scan over the Higgs peak at the muon collider
would then yield an extraordinarily precise determination of $\mh$
along with a determination of the total $\h$ width
that is far more accurate \cite{bbgh1,bbgh2} than achievable by other means 
\cite{summary} in this mass region.
Second, a $\pm 60\mev$ level of accuracy for $\mh$ should prove to
be of great value for constraining parameters entering into
radiative corrections to the Higgs mass.
In the minimal supersymmetric standard model,
the leading one-loop correction to the tree-level prediction for
the mass of the light SM-like $\hl$ is \cite{dpfreport}:
$\Delta\mhl^2=3g^2\mt^4\ln\left(\mstop^2/\mt^2\right)/[8\pi^2\mw^2]$,
where $\mstop$ is the top-squark mass and
we have simplified by neglecting top-squark mixing and non-degeneracy.
From this formula one finds
${d\mhl/ d\mt}\sim 0.6$, and ${d\mhl/ d\mstop}\sim 0.05$,
for $\mhl=100\gev$, $\mt=175\gev$ and
$\mstop\sim 500\gev$. Thus, a $\pm 60\mev$ measurement of
$\mhl$ would translate into very tight constraints
on $\mt$ and $\mstop$ of about $\pm 100\mev$
and $\pm 1.2\gev$, respectively.
Important squark mixing parameters would be similarly constrained.  
The challenge will be to compute higher loop corrections to $\mhl$
to the $\pm 60\mev$ level.

In conclusion, we have shown that with sufficient luminosity it is
possible to determine the Higgs boson mass to a high and
very valuable level of precision by measuring
the $\ell^+\ell^-\to Z\h\to Zb\anti b$ cross section just above
threshold and normalizing to a second measurement either
well above threshold or near the $Z\h$ cross section peak.
One simultaneously determines $\sigma(Z\h)\br(\h\to b\anti b)$ at
a level of accuracy that could distinguish between the Standard Model 
Higgs sector and its many possible (\eg\ supersymmetric) extensions.

\section*{Acknowledgments}

This work was supported in part by the U.S.
Department of Energy
under Grants No. DE-FG02-95ER40896, No.~DE-FG03-91ER40674 and
No.~DE-FG02-91ER40661.
Further support was provided
by the University of Wisconsin Research
Committee, with funds granted by the Wisconsin Alumni Research
Foundation, and by the Davis Institute for High Energy Physics.

\end{document}